\documentclass[showpacs,aps,floatfix,prb,reprint,superscriptaddress]{revtex4-1}
\usepackage{graphicx}
\usepackage{xfrac}
\usepackage{amssymb}
\usepackage{tikz}
\usepackage{amsfonts} %maths
\usepackage{amsmath} %maths
\usepackage[utf8]{inputenc} %useful to type directly diacritic characters
\usepackage{mathtools}
\usepackage{xcolor,colortbl}
\usepackage{bm} %revtex4.1 for bold symbols
\usepackage{array}
\usepackage{color}  % STEFANO
\usepackage{relsize} % for AFLOWpy
 \usepackage[english]{babel}
 \usepackage{amssymb,amsmath}
 \usepackage{color}
 \usepackage{pstricks}
 \usepackage{bbold}
 \usepackage{graphicx}
 \usepackage{color}
 \usepackage{verbatim}
 \usepackage{fancyvrb}
 \usepackage{alltt}
 \usepackage{longtable}
 \usepackage{amsmath}

\definecolor{Blue}{rgb}{0.00,0.00,1.00}
\definecolor{Red}{rgb}{1.00,0.00,0.00}
\definecolor{Green}{rgb}{0.00,0.50,0.00}

\begin{document}

\title{Controlling the TiN electrode work function at the atomistic level: a first principles investigation}

\author{Arrigo Calzolari}
\email {Email: arrigo.calzolari@nano.cnr.it}
\affiliation{CNR-NANO Research Center S3, Via Campi 213/a, 41125 Modena, Italy}
\author{Alessandra Catellani}
\affiliation{CNR-NANO Research Center S3, Via Campi 213/a, 41125 Modena, Italy}

\date{\today}

\begin{abstract}
The paper reports on a theoretical description of work function of TiN, which is one of the most used materials for the realization of electrodes and gates in CMOS devices. Indeed, although the work function is a fundamental quantity in quantum mechanics and also in device physics, as it allows the understanding of band alignment at heterostructures and gap states formation at the metal/semiconductor interface, the role of defects and contaminants is rarely taken into account. Here, by using first principles simulations, we present an extensive study of the work function dependence on nitrogen vacancies and surface oxidation for different TiN surface orientations. The results complement and explain a number of existent experimental data, and provide a useful tool to tailoring transport properties of TiN electrodes in device simulations. 
\end{abstract}

\maketitle

\section{Introduction}
The metal-insulator-metal (MIM) integrated capacitors are key device structures for modern analog and radio frequency  integrated circuits \cite{hu2002}. This includes both two-terminal devices such as in ovonic selectors \cite{chai2019,song2019}, and layered gates in connection with high-$\kappa$ metal-oxide dielectrics, such as, HfO$_2$ \cite{wenger2009}, TiO$_2$\cite{nolan2017}, SiO$_2$\cite{fonseca2006,cottom2019}, and Al$_2$O$_3$\cite{du2018}. 
In this scenario, TiN recently become the reference metal for CMOS compatible gates and MIM devices \cite{lukosius2008}. 
The key feature in the realization of these devices is the alignment between the work function (WF) of the metal contact and the  semiconductor band edges. It is generally assumed that TiN has a WF$\approx 4.7$ eV, even though this value can be
modulated over a large energy range 4.1-5.3 eV, depending on the growth characteristic of the sample, the coupling with semiconductor (e.g. doped Si) or metal-oxide substrates or temperature treatments \cite{fillot2005,liu2006,ren2006,westlinder2003}.

TiN crystallizes in a cubic rocksalt structure, and can be easily cleaved along several low-index faces, such as (100), (110), and (111) surfaces. Previous works indicated a net trend in the surface formation energy of the cleavage surfaces, where TiN(100) is the most and TiN(111) is the least stable one \cite{marlo2000}.
Nonetheless, the growth of single crystal films is unusual and too expensive for any realistic technological application. Almost all TiN-based electrodes and  gate  contacts are made of polycrystalline films, whose constituting grains 
have different sizes and expose multiple faces, depending on the conditions and techniques used to grow the samples. 
In particular, the crystallinity of the substrate strongly affects the morphology, orientation, and resistivity of the films \cite{krylov2019}.
Different sets of XRD characterizations \cite{fillot2005,lukosius2008} indicate that in polycrystalline TiN films grown on SiO$_2$ and HfO$_2$ substrates,  crystalline grains preferentially expose the (111) face, with minor contribution from the (200) one. 
Furthermore, in standard growth conditions (i.e. N-poor) TiN forms stable non-stoichiometric crystals (namely TiN$_x$) over a broad composition range x$\in [0.3 - 1.0]$ \cite{gusev2001,khan1999}. A large variety of multi-technique experiments
\cite{schaffer92,patsalas2001,fink1984,mirguet2006} indicate that in substoichiometric TiN$_x$ materials the most recurrent defects are the nitrogen vacancies (V$_N$), and that high V$_N$ concentrations remarkably affect the optoelectronic \cite{prb_mater,schmid1998} and the transport properties of the system, including its WF\cite{liu2006}. Finally, TiN easily undergoes surface oxidation\cite{milo1995,achour2018}: 
this happens both when as-grown films or fresh cleavage surfaces
are exposed to air \cite{acs_photonics,perrier2014}, and when TiN is in contact with other metal oxides. 

While the role of interfaces with dielectric layers has been thoroughly studied\cite{wenger2009,fonseca2006,chen2011}, the intrinsic structural
and chemical origins of the WF variability have been rarely taken into account \cite{fillot2005}. 
In this paper, we present a first principles investigation of the effects of surface termination, 
substoichiometry and oxidation on TiN WF. Our results indicate that the experimental measured 
WF values are the results of the average combination
of all these structural and composition factors. In particular, the control of defect distribution and crystalline grains 
during growth can
be engineered to tailor the  transport properties of TiN for specific MIM characteristics.

%%%%%%%%%%%%%%%%%%%%%%%%%%%%%%%%%%%%%%%%%%
\section{Computational Details}
%%%%%%%%%%%%%%%%%%%%%%%%%%%%%%%%%%%%%%%%%%
DFT calculations for TiN thin films are performed using the Quantum ESPRESSO package \cite{espresso}. Ab initio ultrasoft  pseudopotentials \cite{uspp} are used to describe the electron-ion interactions and the Perdew-Burke-Ernzerhof (PBE) functional \cite{pbe}, within the generalized gradient approximation (GGA), is used to treat the exchange-correlation electron interaction. In the pseudopotential description, the following valence electron configurations are considered:
N:$2s^22p^3$, O:$2s^22p^4$, and Ti:$2s^22p^63s^23d^2$.
Single particle wavefunction and charge are expanded in a plane wave basis set up to an kinetic energy cutoff of 
30Ry, and 300 Ry, respectively. Geometry optimizations are carried out with convergence thresholds of 0.03eV/\AA ~ for the  the forces on each atom. Extensive accuracy tests can be found in previous publications \cite{prb_TiN,omex,acs_photonics, prb_mater}. 

\begin{figure}
\begin{center}
\includegraphics[width=0.45\textwidth]{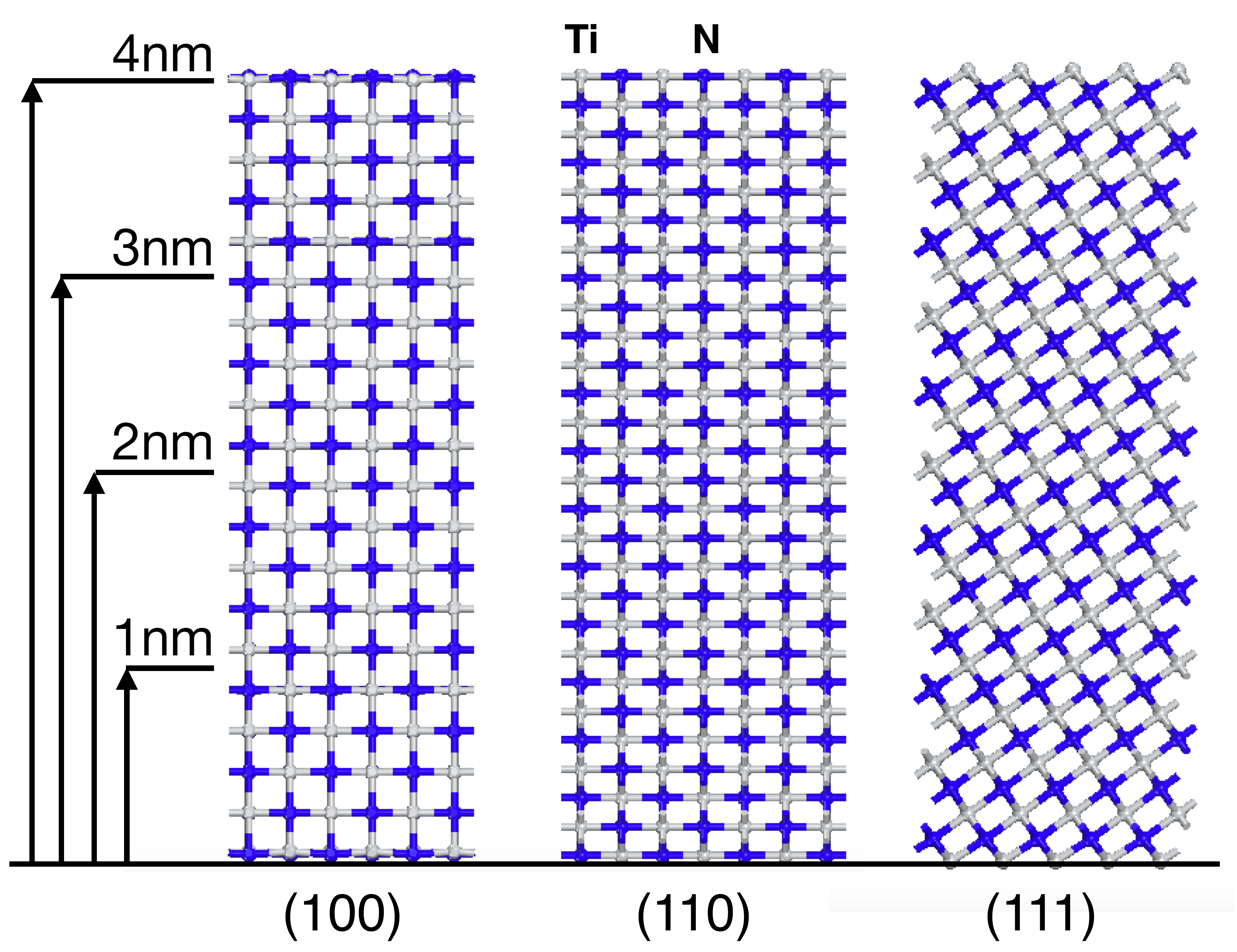}
\caption{Side view atomic structure of TiN surfaces, with (001), (110) and (111) orientation and thickness ranging from
1 nm to 4 nm. }
\label{fig1}
\end{center}
\end{figure}

TiN films are simulated by periodic supercells, where we included a thick vacuum layer ($\sim 15$\AA) in the directions perpendicular to the surface. Each slab contains a variable number of layers with (100), (110) or (111)
surface orientation, which cover the thickness range 1.0-4.0 nm (Figure 1). 
(100) and (110) faces expose an equal number of Ti and N atoms, while (111) surfaces are terminated either with Ti or with N atoms at both slab ends. The layer alternation typical of the [111] stacking of rocksalt structure could impart polar instability to the (111) surface. However, the metallic character of TiN prevents this effects, and no surface reconstruction or non-polar (100) microfacets have been observed in experimental or theoretical reports, confirming that the TiN(111) is a stable surface  \cite{fillot2005,lukosius2008}.
In standard  N-poor growth conditions,
Ti-terminated TiN(111)  surface has the lowest surface energy \cite{yang2020,fan2016}. 
Thus, TiN(111) surface is modeled via a non-polar off-stoichiometric slab with Ti-termination on both external faces (Figure 1c), in agreement with previous theoretical simulations \cite{marlo2000}. No mixed Ti/N terminations of
simulated slabs are considered, in order to avoid spurious
electric fields in the vacuum region of the cell.
N-terminated (111) are not considered in this work, and label (111) will refer only to the Ti-terminated surface. 
Summations over the surface 2D Brillouin zone is done by using dense k-point grids, which depend on the lateral size of the surface, namely ($24\times24$), ($24\times36$), and
($36\times36$), for (100), (110), and (111), respectively. 

Oxide and defective films are obtained starting from a 2 nm-thick TiN reference model, with ($2\times2$) and ($3\times3$)
lateral periodicity for (100) and (111) surface, respectively.  Oxide surfaces are prepared by adding oxygen on 
both outermost layers of TiN slabs (symmetric configuration),  as N-substitutional atoms and O$_2$ adsorbed molecules.  Defective TiN$_x$ systems are simulated including an increasing number of N vacancies (V$_N$) in the reference slabs.

%%%%%%%%%%%%%%%%%%%%%%%%%%%%%%%%%%%%%%%%%%
\section{Results and discussion}
%%%%%%%%%%%%%%%%%%%%%%%%%%%%%%%%%%%%%%%%%%
\subsection{Surface termination}
TiN crystalizes in the NaCl lattice structure within the
\emph{Fm$_3$m} space group; the most favored cleavage surfaces are the TiN(100), TiN(110) and Ti-terminated TiN(111) faces. We studied all these surfaces with different thicknesses from 1nm to 4 nm, as shown in Figure \ref{fig1}. After full atomic optimization, all structures undergo only minor relaxations of the outermost layers, in agreement with previous theoretical calculations \cite{marlo2000,mehmood2015}. Surface relaxation is sufficient to redistribute charge at the surface and to stabilize the structure. The analysis of the projected bandstructures (discussed e.g. in Ref. \cite{prb_TiN}) indicates the presence of surface states in restricted regions (lenses) across the edges of the 2D Brillouin zone at high binding energies, while no surface states are present in the proximity of the Fermi level, for any considered surfaces. 

\begin{figure}
\begin{center}
\includegraphics[width=0.45\textwidth]{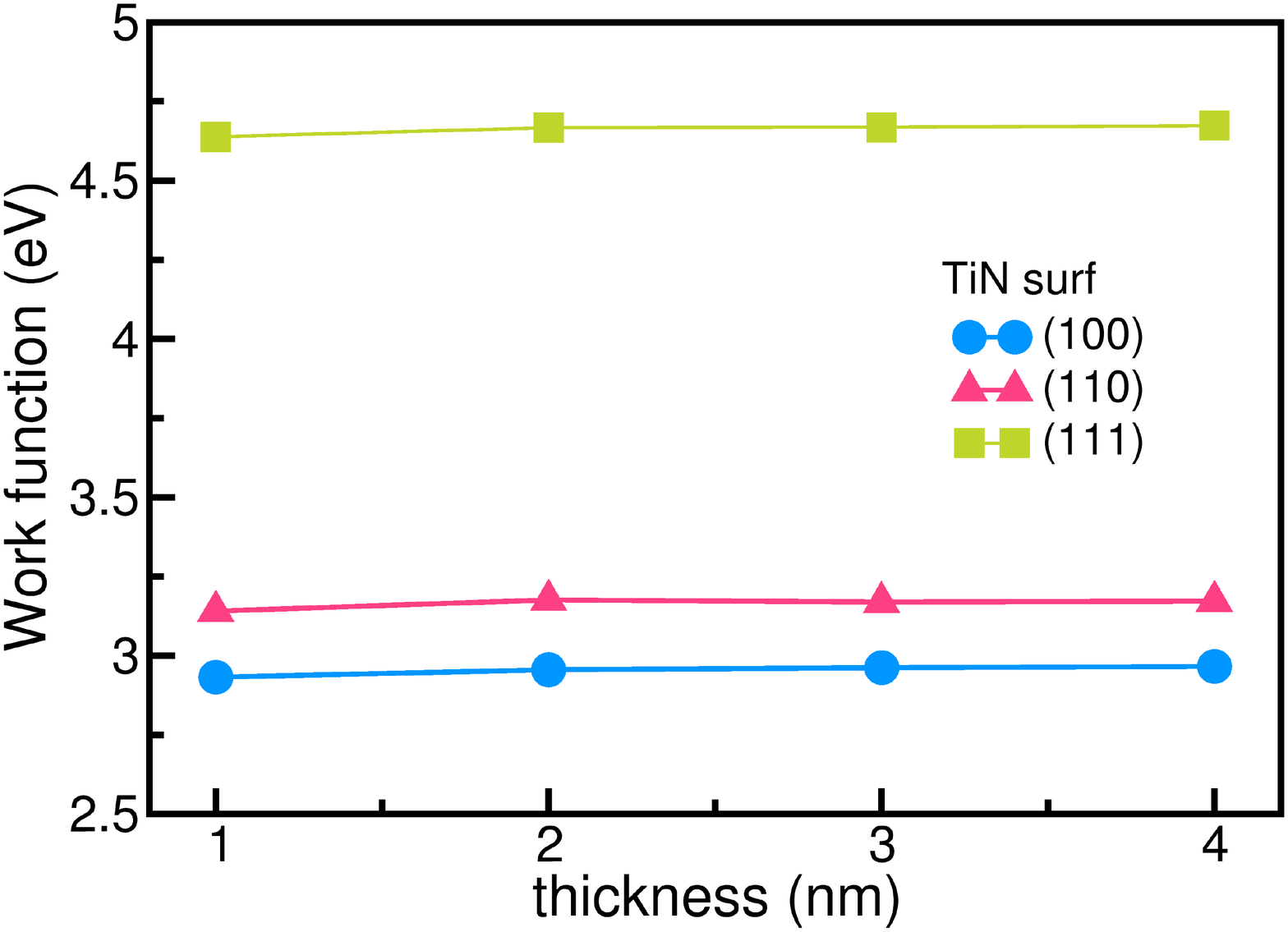}
\caption{Calculated work function for TiN, (100), (110) and (111) surfaces, at variable thickness.}
\label{fig02}
\end{center}
\end{figure}

The work function for all systems is calculated as the difference between the Fermi energy resulting from DFT
calculations and the vacuum level, extracted from the double averaged electrostatic potential $\bar{\bar{V}}_{es}$ \cite{peressi98}.
The results (Figure \ref{fig02}) clearly indicate the effect of the surface termination on the WF of TiN. The calculated WF are 2.96 eV, 3.17 eV and 4.67 eV for (100), (110) and (111) faces, respectively. The WF variability is related to the charge accumulation at surface: (100) and (110) have the same number of Ti and N atoms per layer, while (111) has alternative layer of Ti or N atoms, with a large charge accumulation on the Ti outermost layer.
Thickness hardly affects the WF that, for all surface orientations, remains almost identical with respect to the number of layers ($\Delta$WF=30 meV).  This is due to the the high electron density of TiN ($n_{el}\approx10^{22}$ e/cm$^{-3}$) \cite{prb_TiN}, which easily reaches the bulk-like behavior, in agreement with 
transport and optical measurements on thin TiN films \cite{gall2001,shah2017}. 
These results are well representative of polycrystalline films, where the typical  grain size is of the order of $\sim$2-3 nm \cite{fillot2005}  as well as  of thicker single crystal surfaces \cite{acs_photonics}. 
In particular, the predominance of (111) exposed surfaces in polycrystalline systems \cite{fillot2005} explains the close agreement of the  average experimental  WF values ($\sim 4.6$ eV)  with the calculated results for the (111) surface. 

\subsection{Nitrogen vacancies}
We investigated the effects of V$_N$ in the TiN$_x$(100) and TiN$_x$(111) surface, by
considering an increasing number of N vacancies, from 0 to 50\% of the total nitrogen amount, 
in a reference 2nm-thick TiN(100) and TiN(111) surface, respectively. 
In all cases, the removal of N atoms  from TiN surfaces (i.e. inclusion of N vacancies) does not remarkably change the atomic structure, which maintains the  characteristic of the ideal (undefective) cubic system. 
We do not observe any relevant atomic displacement, clusterization, or sub-phase formation in any system, even for extremely high sub-stoichiometric condition (e.g. removal of 50\% of nitrogen). 
This is in agreement with
the analysis of N-vacancies  in sub-stoichiometric TiN$_x$ bulk, reported in Ref. \cite{prb_mater}. 
The metallic character of TiN prevents the formation of charged defects and charge accumulation around the defects site, which might be responsible for polaronic distortion. 

\begin{figure}
\begin{center}
\includegraphics[width=0.45\textwidth]{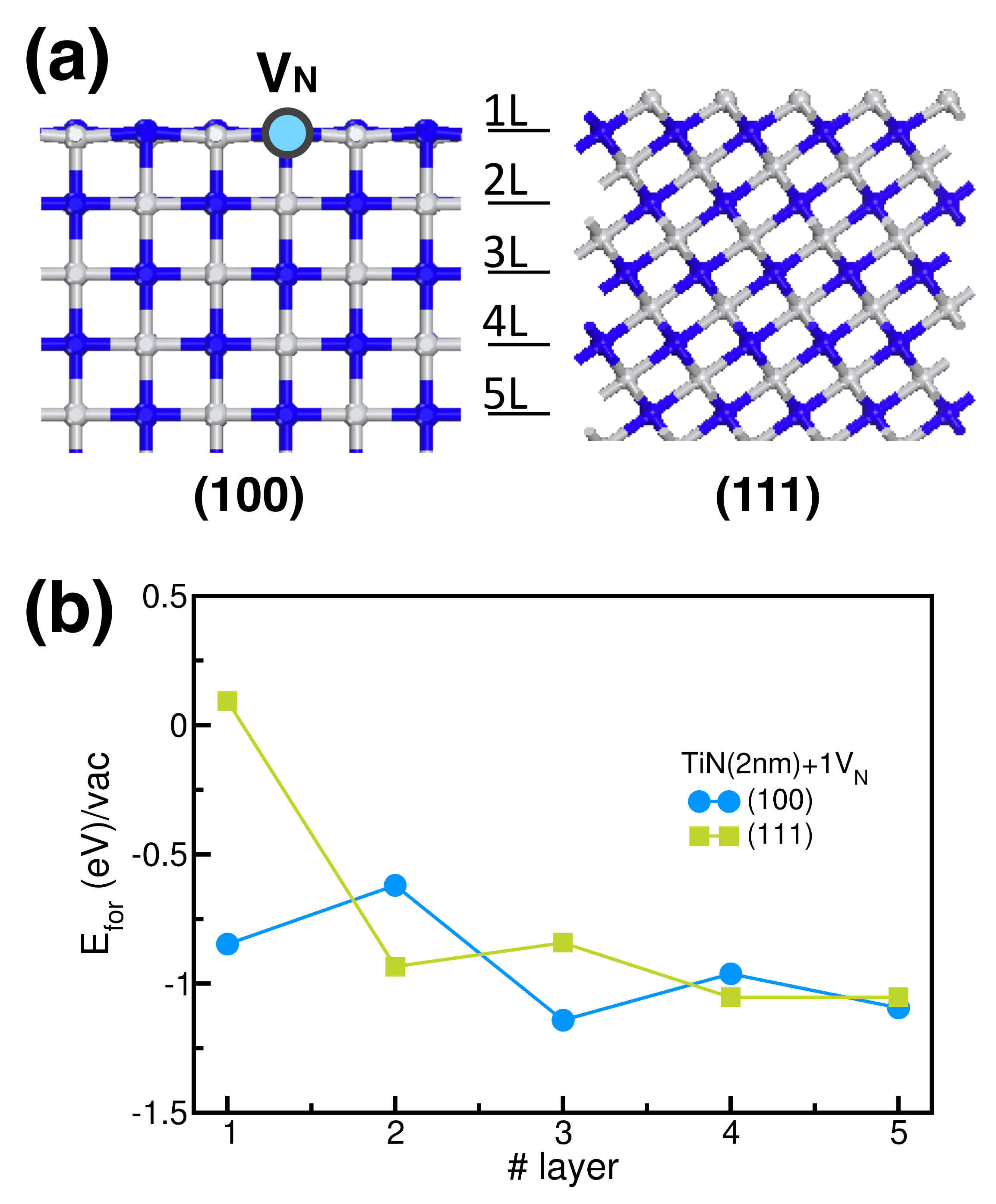}
\caption{(a) Atomic model system for single vacancy V$_N$ (bleu circle) in TiN(100) and TiN(111). 
Labels 1L-5L identify the atomic layer in the slab. (b) Formation energy of single V$_N$, as a function of the atomic layer.}
\label{fig03}
\end{center}
\end{figure}

In order to gain insight on the stability of defective surface systems, we studied the formation energy of a single N vacancy as a function of the layer position.  
As shown in Figure \ref{fig03}a,
we considered the first five layers (1L-5L), where 1L identifies the outermost N-layer, while 5L is representative of an inner
(i.e. bulk like) layer. 
The  formation energy of nitrogen vacancies is defined as \cite{Freysoldt2014}:
\begin{equation}
E_{for}(V_N)_L=E_{tot}(V_N)_L-E_{tot}(surf)+n\mu_N,
\label{eq1}
\end{equation}
where $E_{tot}(V_N)_L$ is the total energy of the optimized TiN$_x$ defective surface including $V_N$ at layer L, $E_{tot}(surf)$ is the total energy of the ideal TiN surface (i.e. no vacancies); $n$  is the number of N-atoms
being removed from a defect-free cell
to its respective reservoir with chemical potential $\mu_N$, to form the defective cell. In the case of a single vacancy, 
$n=1$. Growth conditions determine the bounds limits for the element chemical potential. 
In the Ti-rich/N-poor conditions
N chemical potential in TiN can be obtained as $\mu_{N}=[\mu_{TiN}-\mu_{Ti}]$, where the chemical potential $\mu_{TiN}=E_{tot}(TiN)$ is equal the total energy of TiN bulk (2 atoms per {\em fcc} cell) and $\mu_{Ti}=1/2E_{tot}(Ti^{hcp})$
is the Ti chemical potential extracted from the Ti {\em hcp} metal bulk (2 atoms per cell).
The simulated formation energy of TiN and substoichiometric TiN$_x$ bulk systems have been calculated from first principles in Ref. \cite{prb_mater} in very good agreement with previous results \cite{chase98}.

For TiN(100) the defect formation energy is always negative (Figure \ref{fig03}b), 
with a damped even-odd oscillation which converge in the bulk to the 
value  $E_{for}=$-1.1 eV \cite{prb_mater}, where even layers have the lowest negative formation energies.
TiN(111) also exhibits a damped even-odd trend, but in this case odd  
layers are energetically favored and the first layer has a 
small but positive formation energy. This behavior can be ascribed to the fact that (111) surface is Ti-terminated, 
and N atoms lay in subsurface layers (Figure \ref{fig03}a). 
This general trend confirms that  defective TiN$_x$ surfaces are stable, and energetically favored, with tiny energy differences in the spatial distribution of N-vacancies, which can be considered uniformly distributed over the entire structure. 

\begin{figure}
\begin{center}
\includegraphics[width=0.45\textwidth]{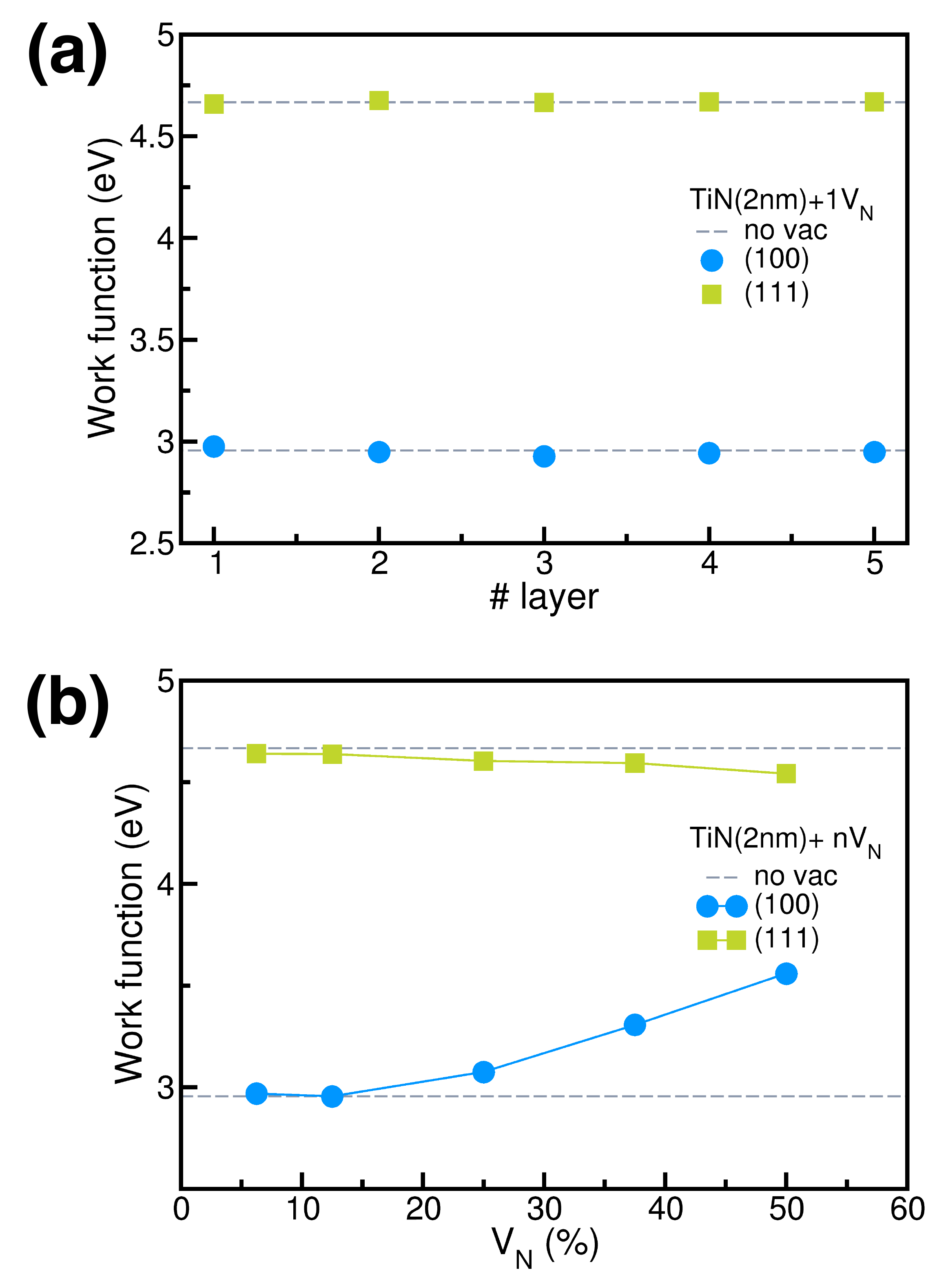}
\caption{WF function of TiN$_x$ in the case of (a) single N vacancy as a function of the layer position, and (b) of multiple 
N vacancies distributed in the slab. The corresponding values for undefective TiN surfaces (dashed lines) are reported for comparison.}
\label{fig04}
\end{center}
\end{figure}

The calculated WFs are shown in Figure \ref{fig04}a, where the corresponding values of undefective TiN surfaces (dashed lines) are  superimposed for comparison. In both cases, the inclusion of a single (i.e. very diluted) N-vacancy does not impart
any relevant change in the WF of the system. However, when the Ti/N ratio increases as in the experimental 
systems, WF deviates from the stoichiometric value,
as displayed in Figure \ref{fig04}b.
The work function of TiN$_x$(100) increases while WF of (111) decreases by 
hundreds of meV, in agreement with the experimental findings \cite{olsson2004}.
As the amount of N content is reduced, the WF of both TiN$_x$ surfaces 
approaches the value WF=4.2 eV, which is a fingerprint also of the {\em hcp} Ti metal. 

This analysis indicates that single WF values, deriving from experiments and used in transport models,
are instead results averaged over the full sample, where the surface terminations and the chemical composition
play a combined role. The statistical predominance of the (111) surface pins the WF final value close to 4.6 eV.
TiN(100) has lower WF, but the presence of N-vacancies shifts WF to higher energy values closer to the (111) 
surface. The final overall value of a polycrystalline electrode depends on
the specific percentage of exposed grain faces and on their  
composition, and thus on the specific growth conditions.
This explains the large variability of the measured WFs. The formation of interfaces with doped semiconductors, of
high-$\kappa$ metal-oxides may further modify these values \cite{fillot2005,lima2012}.

\subsection{Surface oxidation}
\begin{figure}
\begin{center}
\includegraphics[width=0.45\textwidth]{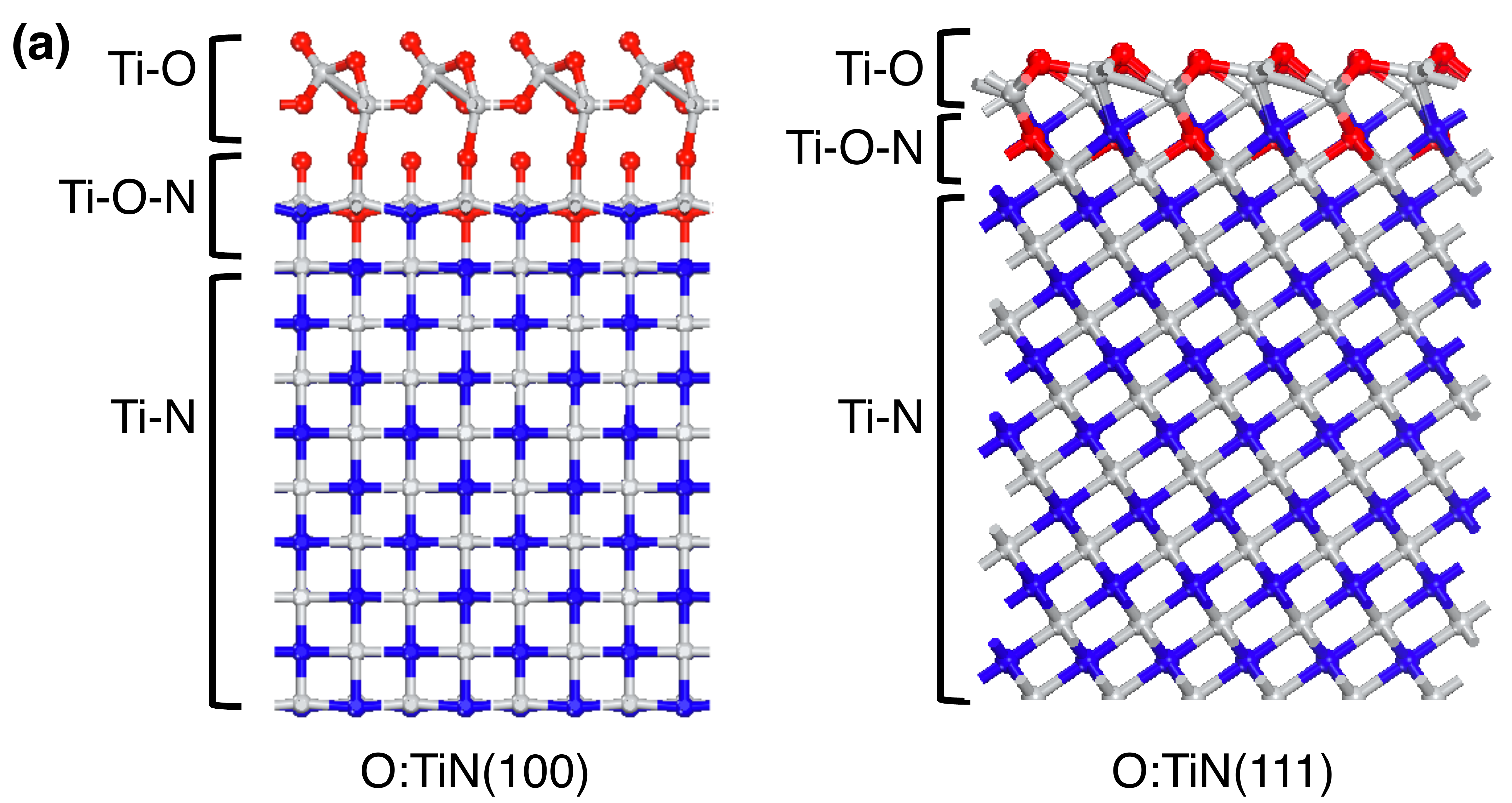}
\includegraphics[width=0.45\textwidth]{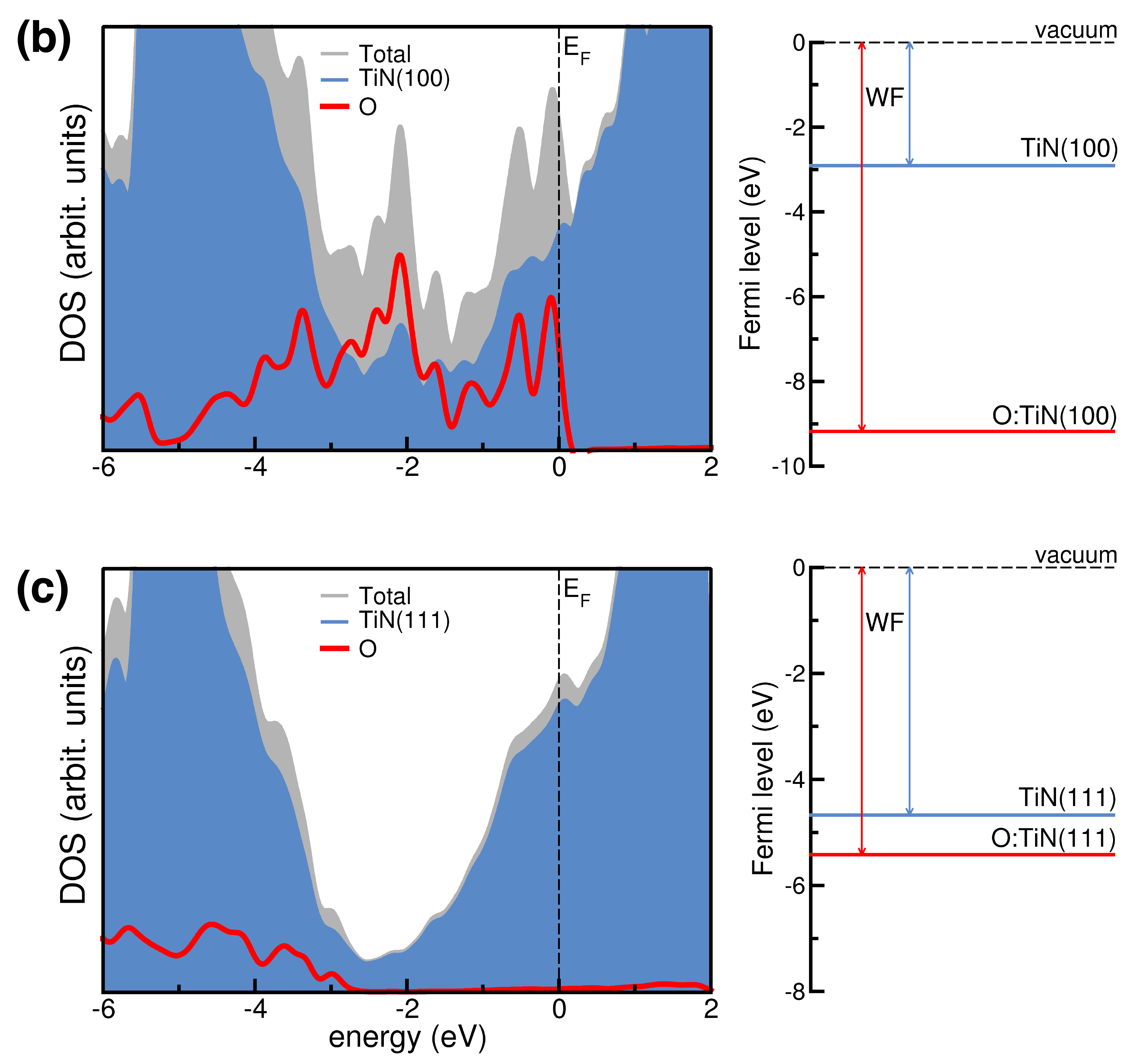}
\caption{(a) Side view of optimized O:TiN(100) and O:TiN(111) structures. Simulation cells have been laterally repeated for clarity. Total and projected DOS (left) and WF
plot (right) for (b) O:TiN(100) and (c) O:TiN(111) structures. Vertical dashed lines in DOS plots indicate the Fermi level, assumed as zero energy reference. In the WF plots (right panels) the zero energy reference is fixed to the vacuum level and the corresponding Fermi levels are shifted accordingly.}
\label{fig05}
\end{center}
\end{figure}

Following the experimental indications\cite{acs_photonics,guler2015} we prepared
the initial O:TiN structures by including a combination of  N-substitutional oxygen atoms and O$_2$ adsorbed molecules on top of 
TiN(100) and TiN(111) surfaces, in both Ti atop and hollow surface sites. 
Different atomic configurations could exist depending on the surface preparation, the growth condition and the oxygen amount. In the case of TiN(100) surface, the initial structure included  3 N-substitutional oxygen atoms and 2 O$_2$ molecules. Other combinations of single O atoms and/or O$_2$ molecules have been previously investigated in Ref. \cite{acs_photonics}. The one presented here best reproduced the transport and optical experimental properties of oxidized ultrathin TiN films \cite{shah2017}. 
The initial O:TiN(111) structure has been prepared along the same line, keeping the same oxygen percentage per TiN atoms of the (100) surface case, and including a mix of N-substitutional and on-surface adsorbed oxygen. 
After atomic optimization, both systems present a strong surface rearrangement and the formation of a mixed oxynitride layer, as shown in Figure \ref{fig05}a. 
Oxygen interacts with TiN surfaces, saturating exposed N-vacancies and bonding to outermost Ti atoms. 
In particular, O$_2$ molecules dissociate to best coordinate with Ti atoms. 
This leads the formation of TiN$_x$O$_y$ layers with different Ti coordinations: 
one intermediate mixed Ti-O-N layer and one Ti-O external capping layer, 
in agreement with X-ray experimental analysis \cite{acs_photonics,milo1995,braic2017}. 
A similar feature is observed also in the XPS oxygen signals where it is  possible to distinguish  N-substitutional, mixed Ti-O-N and pure Ti-O contributions. This implies a different level of coordination of oxygen atoms in the topmost external layers, as shown in Figure \ref{fig05}a. 
Notably, in agreement with experimental results\cite{sia.740230713,zgrabik,braic}, the outermost Ti-O layer does not have the same bonding coordination as in the TiO$_2$ case.  The oxidation of TiN to TiO$_2$ is usually observed at high deposition temperature (T$>$900K) and under constant oxygen fluxes. 

The density of states (DOS) of the resulting systems are shown in Figure \ref{fig05}b-c (left panel).
In the case of (100) surface, the higher amount of under-coordinated oxygen contributes to the states close to the Fermi level (E$_F$), while its contribution
is negligible in the case of (111) surface, where TiN orbitals dominate the energy range close to E$_F$.
This difference is due to the different Ti-coordination in the two surface structures: the coexistence of Ti and N/O in each layer of (100) favors the formation of mixed O-Ti-N bonds. The alternation of Ti and N/O in the (111) structure allows for the formation only of Ti-N and Ti-O bonds at higher binding energies.
The different electronic structures affect also the final WFs of the system that now become 
WF=9.18 eV and WF=5.42 eV, for O:TiN(100) and O:TiN(111), as shown in Figure \ref{fig05}.
Surface oxidation causes a blue-shift of WF in both systems, but while the difference with respect to the clean surface is 
$\Delta$WF=0.75 eV for TiN(111), it is $\Delta$WF=6.30 eV in the case of TiN(100). In the latter case oxygen acts as a 
capping layer that stabilizes the entire system, in agreement with previous theoretical calculations on MgO layers on TiN (100) surface \cite{ren2006}. The further growth of thick metal-oxide layers may mitigate this effect
resulting in WF closer to range 4-5 eV, typically assumed for TiN/oxide gate interfaces \cite{fillot2005,liu2006,ren2006}. 
Thus, the total or partial oxidation of crystal grains constitutes another degree of freedom in 
the overall WF value of  TiN electrodes. 

\section{Conclusion}
We presented a detailed study of the structural, stoichiometric and chemical effects on the work function
of intrinsic TiN films, as used in electrode contacts. Our results show that face orientation, N-vacancies and surface
oxidations are different causes of modulation of TiN WF, whose final value depends on the details of the
growth conditions. The typical predominance of TiN$_x$ grains with (111) terminations,  
is the main origin of the average WF measured  in polycrystalline electrodes.
The control of the growth conditions can be engineered  to obtain a fine tuning of the electrode WF 
so to optimize the band alignment with the semiconductor elements of the devices. 

\section{Acknowledgments}
This work was supported in part by EC through H2020-NMBP-TO-IND project GA n. 814487 (INTERSECT).


\begin{thebibliography}{00}

\bibitem{hu2002} H. Hu, C. Zhu, Y. F. Lu, M. F. Li, B.J . Cho, W. K. Choi,``A high performance MIM capacitor using HfO$_2$ dielectrics,'' \emph{IEEE Electron. Device Lett.} vol. 23, pp. 514--516, Oct. 2002.

\bibitem{chai2019} Z. Chai, W. Zhang, R. Degraeve, S. Clima, F. Hatem, J. F. Zhang, P. Freitas, J. Marsland, A. Fantini, D. Garbin, L. Goux, and G. Sankar Kar,``Dependence of Switching Probability on Operation Conditions in Ge$_x$Se$_{1?x}$ Ovonic Threshold Switching Selectors,'' \emph{IEEE Electron. Device Lett.} vol. 40, pp. 1269--1272, Aug. 2019.

\bibitem{song2019} B. Song, H. Xu, S. Liu, H. Liu, Q. Liu, and Q. Li,``An ovonic threshold switching selector based on Se?rich GeSe chalcogenide,'' \emph{Appl. Phys. A} vol. 125, pp. 772, Oct. 2019.

\bibitem{wenger2009} Ch. Wenger, M. Lukosius, G. Weidner, H.-J. M\"ussig, S. Pasko, and Ch. Lohe, ``The role of the HfO$_2$-TiN interface in capacitance-voltage nonlinearity of Metal-Insulator-Metal capacitors,'' \emph{Electroch. Acta} vol. 141, pp. 25--32, Jul. 2014.

\bibitem{nolan2017} J. J. Guti\'errez Moreno, and M. Nolan, ``Ab Initio Study of the Atomic Level Structure of the Rutile TiO2(110)? Titanium Nitride (TiN) Interface,'' \emph{ACS Appl. Mater. Inter.} vol. 9, pp. 38089--38100, Oct. 2017.

\bibitem{fonseca2006} L. R. C. Fonseca and A. A. Knizhnik, ``First-principles calculation of the TiN effective work function on SiO$_2$ and on HfO$_2$,'' \emph{Phys. Rev. B} vol. 74, pp. 195304, Nov. 2006.

\bibitem{cottom2019} J. Cottom, A. Bochkarev, E. Olsson, K. Patel, M. Munde, J. Spitaler, M. N. Popov, M. Bosman, and A. L. Shluger, ``Modeling of Diffusion and Incorporation of Interstitial Oxygen Ions at the TiN/SiO$_2$ Interface,'' \emph{ACS Appl. Mater. Inter.} vol. 11, pp. 36232--36243, Sep. 2019.

\bibitem{du2018} L. Du, H. Wong, S. Dong, W.-S. Lau, and V. Filip, ``AFM study on the surface morphologies of TiN films prepared by magnetron sputtering and Al$_2$O$_3$ films prepared by atomic layer deposition,'' \emph{Vacuum} vol. 153, pp. 139--144, Apr. 2018.

\bibitem{lukosius2008} M. Lukosius, C. Wenger, S. Pasko, H.-J. M\"ussig, B. Seitzinger, and C. Lohe, ``Atomic Vapor Deposition of Titanium Nitride as Metal Electrodes for Gate-last CMOS and MIM Devices,'' \emph{Chem. Vapor Dep.} vol. 14, pp. 123--128, Jun. 2008.

\bibitem{fillot2005} F. Fillot, T. Morel, S. Minoret, I. Matko, S. Maitrejean, B. Guillaumot, B. Chenevier, and T. Billon, ``Investigations of titanium nitride as metal gate material, elaborated by metal organic atomic layer deposition using TDMAT and NH$_3$,'' \emph{Microelect. Eng.} vol. 82, pp. 248--253, Dec. 2005.

\bibitem{liu2006}Y. Liu, S. Kijima, E. Sugimata, M. Masahara, K. Endo, T. Matsukawa, K. Ishii, K. Sakamoto, T. Sekigawa, H. Yamauchi, Y. Takanashi, and Eiichi Suzuki, ``Investigation of the TiN Gate Electrode
With Tunable Work Function and Its Application for FinFET Fabrication,'' \emph{IEEE Trans.  Nanotech.} vol. 5, pp. 723--730, Nov. 2006.

\bibitem{ren2006}C. Ren, B. B. Faizhal, D.S.H. Chan, M.-F. Li, Y.-C. Yeo, A.D. Trigg, N. Balasubramanian, D.-L. Kwong, ``Work function tuning of metal nitride electrodes for advanced CMOS devices,'' \emph{Thin Sol. Films} vol. 505, pp. 174--177, May. 2006.

\bibitem{westlinder2003}J. Westlinder, T.  Schram, L.  Pantisano, E. Cartier, A.  Kerber, G. S.  Lujan, J.  Olsson, and G. Groeseneken, ``On the thermal stability of atomic layer deposited TiN as gate electrode in MOS devices,'' \emph{IEEE Electron. Device Lett.} vol. 24, pp. 550--552, Sep. 2003.

\bibitem{marlo2000} M. Marlo and V. Milman, ``Density-functional study of bulk and surface properties of titanium nitride using different exchange-correlation functionals,'' \emph{Phys. Rev. B} vol. 62, pp. 2899--2907, Jul. 2000.

\bibitem{krylov2019} I. Krylov, X. Xu, Y. Qi, K. Weinfeld, V. Korchnoy, M. Eizenberg, and D. Ritter, ``Effect of the substrate on structure and properties of titanium nitride films grown by plasma enhanced atomic layer deposition,'' \emph{J. Vac. Sci. Technol. A} vol. 37, pp. 060905, Sep. 2019.

\bibitem{gusev2001}A. I. Gusev, A. A. Rempel, and A. J. Magerl '' \emph{Disorder and Order in Strongly Nonstoichiometric Compounds}  Springer, Heidelberg 2001.

\bibitem{khan1999}J. H. Kang, and K. J. Kim, ``Structural, optical, and electronic properties of cubic TiNx
TiN$_x$ compounds,'' \emph{J. Appl. Phys.} vol. 86, pp. 346--350, Jun. 1999.

\bibitem{schaffer92} J. P. Schaffer, A. J. Perry, and J. Brunner, ``Defects in hard coatings studied by positron annihilation spectroscopy and x?ray diffraction,'' \emph{J. Vac. Sci. Technol. A} vol. 10, pp. 193--207, Jan. 1992.

\bibitem{patsalas2001} P. Patsalas, and S. Logothetidisr, ``Optical, electronic, and transport properties of nanocrystalline titanium nitride thin films,'' \emph{J. Appl. Phys.} vol. 90, pp. 4725--4734, Nov. 2001.

\bibitem{fink1984} J. Pfl\"uger, J. Fink, W. Weber, K. P.  Bohnen, and G. Crecelius, ``Dielectric properties of ${\mathrm{TiC}}_{x}$, ${\mathrm{TiN}}_{x}$, ${\mathrm{VC}}_{x}$, and ${\mathrm{VN}}_{x}$ from 1.5 to 40 eV determined by electron-energy-loss spectroscopy,'' \emph{Phys. Rev. B} vol. 30, pp. 1155--1163, Aug. 1984.

\bibitem{mirguet2006} C. Mirguet, L. Calmels, and Y. Kihn, ``Electron energy loss spectra near structural defects in TiN and TiC,'' \emph{Micron} vol. 37, pp. 442--448, Jul. 2006.

\bibitem{prb_mater} A. Catellani, P. D'Amico,  and A. Calzolari, ``Tailoring the plasmonic properties of metals: The case of substoichiometric titanium nitride,'' \emph{Phys. Rev. Mater.} vol. 4, pp. 015201, Jan. 2020.

\bibitem{schmid1998} P. E. Schmid, M. S. Sunaga, and F. L\'evy, ``Optical and electronic properties of sputtered thin films,'' \emph{J. Vac. Sci. Technol. A} vol. 16, pp. 2870--2875, May. 1998.

\bibitem{milo1995} I. Milo\v sv,  H.-H. Strehblow,  B. Navin\v sek, and  M. Metiko\v s?Hukovi\'c, ``Electrochemical and thermal oxidation of TiN coatings studied by XPS,'' \emph{Surf. Interf. Analysis} vol. 23, pp. 529--539, Jul. 1995.

\bibitem{achour2018} A. Achour, M. Islam, I. Ahmad, L. Le Brizoual, A. Djouadi, and T. Brousse, ``Influence of surface chemistry and point defects in TiN based electrodes on electrochemical capacitive storage activity,'' \emph{Scripta Materialia} vol. 153, pp. 59--62, May. 2018.

\bibitem{acs_photonics} D. Shah, A. Catellani, H. Reddy, N. Kinsey, V. Shalaev, A. Boltasseva, and A. Calzolari, ``Controlling the Plasmonic Properties of Ultrathin TiN Films at the Atomic Level,'' \emph{ACS Photonics} vol. 5, pp. 2816--2824, May 2018.

\bibitem{perrier2014} A. Seifitokaldania, O. Savadogo, and M. Perrier, ``Density Functional Theory (DFT) Computation of the Oxygen Reduction Reaction (ORR) on Titanium Nitride (TiN) Surface,'' \emph{This Sol. Films} vol. 517, pp. 6334--6336, Feb. 2009.

\bibitem{chen2011} C.-L. Chen, and Y.-C. King, ``TiN Metal Gate Electrode Thickness Effect
on BTI and Dielectric Breakdown in HfSiON-Based MOSFETs,'' \emph{IEEE Trans. Elecron. Dev} vol. 58, pp. 3736--3742, Nov. 2011.

\bibitem{espresso} P. Giannozzi, O. Andreussi, T. Brumme, O. Bunau, M. Buongiorno Nardelli, M. Calandra, R. Car, C. Cavazzoni, D. Ceresoli, M. Cococcioni, N. Colonna, I. Carnimeo, A. D. Corso, S. de Gironcoli, P. Delugas, R. A. DiStasio, A. Ferretti, A. Floris, G. Fratesi, G. Fugallo, R. Gebauer, U. Gerstmann, F. Giustino, T. Gorni, J. Jia, M. Kawamura, H.-Y. Ko, A. Kokalj, E. K\:uckbenli, M. Lazzeri, M. Marsili, N. Marzari, F. Mauri, N. L. Nguyen, H.-V. Nguyen, A. O. de-la Roza, L. Paulatto, S. Ponc\'e, D. Rocca, R. Sabatini, B. Santra, M. Schlipf, A. P. Seitsonen, A. Smogunov, I. Timrov, T. Thonhauser, P. Umari, N. Vast, X. Wu, and S. Baroni, ``Advanced capabilities for materials modelling with Quantum ESPRESSO,'' \emph{J. Phys.: Condens. Matter} vol. 29, pp. 465901, Oct. 2017.

\bibitem{uspp} D. Vanderbilt,``Soft self-consistent pseudopotentials in a generalized eigenvalue
         formalism,'' \emph{Phys. Rev. B} vol. 41, pp. R7892--R7895, Apr. 1990.

\bibitem{pbe} J. P. Perdew, K. Burke, and M. Ernzerhof, ``Generalized Gradient Approximation Made Simple,'' \emph{Phys. Rev. Lett.} vol. 77, pp. 3865--3868, Oct. 1996.

\bibitem{prb_TiN} A. Catellani and A. Calzolari, ``Plasmonic properties of refractory titanium nitride,'' \emph{Phys. Rev. B} vol. 95, pp. 115145, Mar. 2017.

\bibitem{omex}A. Catellani and A. Calzolari, ``Tailoring the plasmonic properties of ultrathin TiN films at metal-dielectric interfaces,'' \emph{Opt. Mater. Exp.} vol. 9 pp. 1458--1458, Mar. 2019.

\bibitem{yang2020} T. Yang, M. Wei, Z. Ding, X. Han, and J. Li, ``Ab initio calculations on the Mg/TiN heterogeneous nucleation interface,'' \emph{J. Phys. Chem. Sol.} vol. 143, pp. 109479, Apr. 2020.

\bibitem{fan2016} X. Fan, B. Chen, M. Zhang, D. Li, Z. Liu, and C. Xiao, ``First-principles calculations on bonding characteristic and electronic property of TiC (111)/TiN (111) interface,'' \emph{Mater. \& Design.} vol. 112, pp. 282--289, Dec. 2016.

\bibitem{mehmood2015} F. Mehmood, R. Pachter, N. R. Murphy, and W. E. Johnson, ``Electronic and optical properties of titanium nitride bulk and surfaces from first principles calculations,'' \emph{J. Appl. Phys.} vol. 118, pp. 195302, Nov. 2015.

\bibitem{peressi98} M Peressi and N Binggeli and A Baldereschi, ``Band engineering at interfaces: theory and numerical experiments,'' \emph{J. Phys.D: Appl. Phys.} vol. 31, pp. 1273--1299, Jun. 1998.

\bibitem{shah2017} D. Shah, H. Reddy, N. Kinsey, V. M. Shalaev, and A. Boltasseva, ``Optical Properties of Plasmonic Ultrathin TiN Films,'' \emph{Adv. Opt. Mater.} vol. 5, pp. 1700065, May 2017.

\bibitem{gall2001} D. Gall, I.Petrov, and J. E. Greene, ``Epitaxial Sc$_{1?x}$Ti$_x$N(001): Optical and Electronic Transport Properties,'' \emph{J. Appl. Phys.} vol. 89, pp. 401--409, Jan. 2001.

\bibitem{Freysoldt2014} C. Freysoldt, B. Grabowski, T. Hickel, J. Neugebauer, G. Kresse, A. Janotti and C. G. Van de Walle,``First-principles calculations for point defects in solids,'' \emph{Micron} vol. 86, pp. 253--304, Jan.-Mar. 2014.

\bibitem{chase98} M. W. Chase Jr.``First-principles calculations for point defects in solids,
\emph{J. Phys. Chem. Ref. Data, Monograph} vol. 9, 1--1951 1998.

\bibitem{olsson2004} J. Westlinder, G. Sj\"oblom, and J. Olsson, ``Variable work function in MOS capacitors utilizing nitrogen-controlled TiN$_x$ gate electrodes,'' \emph{Microelect. Eng.} vol. 75, pp. 389--396, Aug. 2004.

\bibitem{lima2012} L. P. B. Lima, J. A. Diniz, I. Doi, and J. Godoy Fo, ``Titanium nitride as electrode for MOS technology and Schottky diode: Alternative extraction method of titanium nitride work function,'' \emph{Microelect. Eng.} vol. 92, pp. 86--90, Apr. 2012.

\bibitem{guler2015} U. Guler, S. Suslov, A. V. Kildishev, A. Boltasseva, and V. M. Shalaev,  ``Colloidal Plasmonic Titanium Nitride Nanoparticles: Properties and Applications,'' \emph{Nanophotonics} vol. 4, pp. 269--276, Jun 2015.

\bibitem{braic2017} L. Braic, N. Vasilantonakis, A. Mihai, I. J. Villar Garcia, S. Fearn, B. Zou, N. McN. Alford, B. Doiron, R. F. Oulton, S. A. Maier, A. V. Zayats, and P. K. Petrov, ``Titanium Oxynitride Thin Films with Tunable Double Epsilon-Near-Zero Behavior for Nanophotonic Applications,'' \emph{ACS Appl. Mater. Inter.} vol. 9, pp. 29857--29862, Aug 2017.

\bibitem{sia.740230713} I. Milos\v v, H.-H. Strehblow, B. Navins\v ek, and M. Metiko\v s-Hukovi\'c. ``Electrochemical and Thermal Oxidation of TiN Coatings Studied by XPS,''  \emph{Surf. Interface Anal.} vol. 23,  pp. 529--539, Jul 1995.

\bibitem{zgrabik} C. M. Zgrabik, and E. L.  Hu, ``Optimization of Sputtered Titanium Nitride as a Tunable Metal for Plasmonic Applications,''  \emph{Opt. Mater. Express} vol. 5, pp. 2786--2797, Nov. 2015.

\bibitem{braic} L. Braic, N. Vasilantonakis, A. Mihai, I. J.  Villar Garcia, S. Fearn, B. Zou, N. M.  Alford, B. Doiron, R. F. Oulton, S. A. Maier, A. V. Zayats, and P. K. Petrov, ``Titanium Oxynitride Thin Films with Tunable Double Epsilon-Near- Zero Behavior for Nanophotonic Applications,''  \emph{ACS Appl. Mater. Interfaces} vol. 9 , pp. 29857--29862, Aug. 2017.

\end{thebibliography}
\end{document}